\documentstyle[12pt,a4,a4wide,epsfig]{article}
\def\Journal#1#2#3#4{{#1} {\bf #2}, #3 (#4)}

\def\NPBP{{\em Nucl. Phys.} {\bf B} (Proc. Suppl.)}

\def\PLB{{\em Phys. Lett.}  {\bf B}}
\def\PRL{\em Phys. Rev. Lett.}
\def\PRD{{\em Phys. Rev.} {\bf D}}
\def\PRC{{\em Phys. Rev.} {\bf C}}

\def\PRP{{\em Phys. Rep. }}
\def\EPC{{\em Eur. Phys. J.} {\bf C}}
\def\ZPC{{\em Z. Phys.} C}

\def\PNPP{\em Prog. Nucl. Part. Phys.}
\def\PTP{\em Prog. Theo. Phys.}

\def\APJ{\em Astrophys. J.}

\def\RMP{\em Rev. Mod. Phys.}

\def\EPL{\em Europhys. Lett.}

\def\MPL{{\em Mod. Phys. Lett.} {\bf A}}

\def\JHEP{\em JHEP}

\def\ra{\rightarrow}
\def\be{\begin{equation}}
\def\ee{\end{equation}}

\def\gs{\mathrel{%
   \rlap{\raise 0.511ex \hbox{$>$}}{\lower 0.511ex \hbox{$\sim$}}}}
\def\ls{\mathrel{
   \rlap{\raise 0.511ex \hbox{$<$}}{\lower 0.511ex \hbox{$\sim$}}}}

\newcommand{\obb}{0\mbox{$\nu\beta\beta$}}

\newcommand{\onbb}{neutrinoless double beta decay}

\newcommand{\uee}{\mbox{$|U_{e1}|^2$}}
\newcommand{\uez}{\mbox{$|U_{e2}|^2$}}
\newcommand{\ued}{\mbox{$|U_{e3}|^2$}}
\newcommand{\ume}{\mbox{$|U_{\mu 1}|^2$}}

\newcommand{\umd}{\mbox{$|U_{\mu 3}|^2$}}
\newcommand{\ute}{\mbox{$|U_{\tau 1}|^2$}}

\newcommand{\utd}{\mbox{$|U_{\tau 3}|^2$}}

\newcommand{\delm}{\mbox{$\Delta m^2$} }

\newcommand{\sint}{\mbox{$\sin^2 2\theta$}}

\newcommand{\cs}{cross section}

\newcommand{\ba}{\begin{array}{c}}
\newcommand{\baz}{\begin{array}{cc}}
\newcommand{\bad}{\begin{array}{ccc}}
\newcommand{\bea}{\begin{equation} \begin{array}{c}}
\newcommand{\eea}{ \end{array} \end{equation}}
\newcommand{\ea}{\end{array}}

\newcommand{\aen}{\mbox{$\overline{\nu_e} $}}

\newcommand{\nnu}{\mbox{$0\nu \beta \beta$}}

\newcommand{\mab}{\mbox{$\langle m_{\alpha \beta} \rangle $}}
\newcommand{\mmm}{\mbox{$\langle m_{\mu \mu} \rangle$}}
\newcommand{\mtta}{\mbox{$\langle m_{\tau \tau} \rangle$}}
\newcommand{\mmt}{\mbox{$\langle m_{\mu \tau} \rangle$}}
\newcommand{\mee}{\mbox{$\langle m_{ee} \rangle$}}
\newcommand{\met}{\mbox{$\langle m_{e \tau} \rangle$}}
\newcommand{\meu}{\mbox{$\langle m_{e \mu} \rangle$}}

\begin{document}
\title{\hfill { \normalsize DO--TH 00/02}\\
\hfill {\normalsize hep-ph/0003149}\\\vskip 1.5cm
\bf Neutrino oscillation experiments and limits on lepton--number 
and lepton--flavor violating processes}
\author{Werner Rodejohann\footnote{Email address: rodejoha@dilbert.physik.uni-dortmund.de}\\
{\it \normalsize Lehrstuhl f\"ur Theoretische Physik III,}\\ 
{\it \normalsize Universit\"at Dortmund, Otto--Hahn Str. 4,}\\ 
{\it \normalsize 44221 Dortmund, Germany}}
\date{}
\maketitle
\thispagestyle{empty}
\begin{abstract}
Using a three neutrino framework we investigate bounds for 
the effective Majorana neutrino mass matrix. 
The mass measured in neutrinoless double beta decay 
is its (11) element. 
Lepton--number and --flavor violating processes sensitive to 
each element are considered and limits on branching ratios or cross 
sections are given. 
Those processes include $\mu^- e^+$ conversion, $K^+ \to \pi^- \mu^+ \mu^+$ 
or recently proposed high--energy scattering processes at HERA. 
Including all possible mass schemes, the three solar solutions and other 
allowed possibilities, there is a total of 80 mass matrices. 
The obtained indirect limits are up to 14 orders of magnitude 
more stringent than direct ones.  
It is investigated how neutrinoless double beta decay may judge between 
different mass and mixing schemes as well as 
solar solutions. 
Prospects of detecting processes depending on elements of the 
mass matrix are also discussed.     
\end{abstract}

\newpage
\section{Introduction}
In recent years overwhelming evidence for non--vanishing neutrino 
masses was collected in atmospheric \cite{SK,atmo},
solar \cite{solar} and accelerator \cite{LSND} experiments, 
opening a window into a variety of new phenomena \cite{kaireport}. 
The results are interpreted in terms of neutrino oscillations 
governed by a mixing angle and a mass--squared difference. 
Analyzing the data of the respective experiments yields typically
values of 
\be \label{results}
\bad 
(\delm  \; {\rm (eV^2)},\; \sint) \simeq & \left\{ \hspace{-2cm} 
\begin{array}{cc} 
(10^{-3} , \gs 0.7) & \mbox{ atmospheric} \\[0.3cm]
\left.
\begin{array}{cc}
\hspace{2cm} (10^{-5} , 10^{-3})  & \mbox{ SAMSW} \\[0.2cm]
\hspace{2.1cm} (10^{-5} , \gs 0.7)  & \mbox{ LAMSW} \\[0.2cm]
\hspace{2.2cm} (10^{-10} , \gs 0.7) & \mbox{ VO}   \\[0.2cm]
\end{array}  \right\} & \mbox{ solar} \\[0.3cm] 
(1 ,  10^{-3}) & \mbox{ LSND.} \end{array} \right.
\ea
\ee
Here SAMSW (LAMSW) denotes the small (large) angle MSW \cite{MSW} solution  
and VO the vacuum oscillation solution of the solar neutrino problem. 
One sees that the atmospheric and solar mass scale obey 
the following relation: 
\be \label{scales}
\Delta m_{\odot}^2 \ll \Delta m_{\rm A}^2  ,  
\ee
regardless of the solar solution chosen. 
In order to avoid sterile neutrinos one usually  
leaves out the LSND \cite{LSND} result (which also 
gives time to wait until the conflict with KARMEN 
\cite{KARMEN} is resolved). We are thus working in a 
three neutrino framework, which allows a simple derivation of the 
leptonic Maki--Nakagawa--Sakata (MNS) matrix \cite{MNS}. 
Only five mass schemes can accommodate the results in a 
three neutrino framework, 
three hierarchical and two degenerate schemes with the overall scale 
given by cosmological arguments to be of the order of 
a few eV\@. These mass schemes can be divided into two scenarios. 
In addition, we distinguish the cases in which one element of 
the mixing element is zero as indicated by CHOOZ data and the general case 
of all elements being non--zero. Ordering effects of 
some elements are also considered.\\  
From the MNS matrix one can infer limits on the $3 \times 3$ matrix of 
effective Majorana neutrino masses defined as 
\bea \label{defin}
\mab = |(U \, {\rm diag}(m_1 , m_2 , m_3) U^{\rm T})_{\alpha \beta}| \\[0.3cm] 
= \left| \sum m_i U_{\alpha i} U_{\beta i} \right| 
\le \sum m_i | U_{\alpha i} U_{\beta i}| 
\mbox{ with }  \alpha, \beta = e , \, \mu  , \, \tau , 
\eea
where $U$ is the mixing matrix and the $m_i$ are mass eigenvalues. With 
this approximation, \mab{} is symmetrical. 
The mass measured in \onbb{} ($\obb$) 
is the $(ee)$ element of this matrix and has been 
considered by several authors \cite{mee1,mee2}. 
However, the complete matrix is rarely analyzed in terms of 
phenomenological consequences, except for 
the elements \meu{} and \mmm{} in Refs.\ \cite{Japaner1}, 
although without giving concrete numbers. 
We introduce processes dependent on every element of \mab{}, 
like $\mu^- e^+$ conversion, $K^+ \to \pi^- \mu^+ \mu^+$ or recently 
proposed  
high--energy scattering processes, finding rather 
discouraging results for branching ratios or life times. 
We compare our results (calling them {\it indirect} limits)
with current experimental ({\it direct}) bounds  
and find that our limits are up to 14 orders of 
magnitude more stringent than current experimental data and therefore 
beyond experimental access in the near future. 
Special attention is paid to \obb{} and its sensitivity on 
mass and mixing. Though it will be much 
easier to wait for further astrophysical data to distinguish 
between vacuum and MSW solutions for the solar neutrino problem, 
it would be a remarkable 
experiment to decide via terrestrial 
nuclear physics experiments if matter effects inside the Sun 
are of importance or not. 
Unfortunately, it turns out that the only scheme delivering  
values inside the range of next generation experiments is 
insensitive on the solar solutions.\\
The paper is organized as follows: 
In Section 2 we present the usual assumptions that lead to the derivation of 
the MNS matrix and \mab{}. 
Section 3 gives the results and pays special attention to the 
$(ee)$ element of \mab{}. In Section 4 processes are introduced which are 
sensitive on the respective elements of \mab{} just as \obb{} is 
on \mee{}. Section 5 concludes the paper.

\section{Oscillation probabilities and mass schemes}
\begin{figure}[t]
\begin{center}
\begin{picture}(8.6,6)
\put(0,0){\line(1,0){2.1}}
\put(0,1.5){\line(1,0){2.1}}
\put(0,4){\line(1,0){2.1}}
\put(2.9,0){\line(1,0){2.1}}
\put(2.9,0.2){\line(1,0){2.1}}
\put(2.9,4){\line(1,0){2.1}}
\put(5.8,2){\line(1,0){2.1}}
\put(5.8,2.2){\line(1,0){2.1}}
\put(5.8,1.9){\line(1,0){2.1}}
\put(0.8,0.2){$m_1$}
\put(3.2,0.4){$m_1 \simeq m_2$}
\put(0.8,1.7){$m_2$}
\put(0.8,4.2){$m_3$}
\put(3.7,4.2){$m_3$}
\put(8,2.0){$m_0$}
\put(0,5){\bf {\small Scheme A I}}
\put(2.9,5){\bf {\small Scheme A II}}
\put(5.72,5){\bf {\small Scheme A III}}
\end{picture}
\vspace{1cm}
\caption{\label{massschemes1}The three mass schemes that can accommodate the 
relation $\Delta m^2_{21} \ll \Delta m^2_{31} \simeq \Delta m^2_{32}$.}
\end{center}
\end{figure}
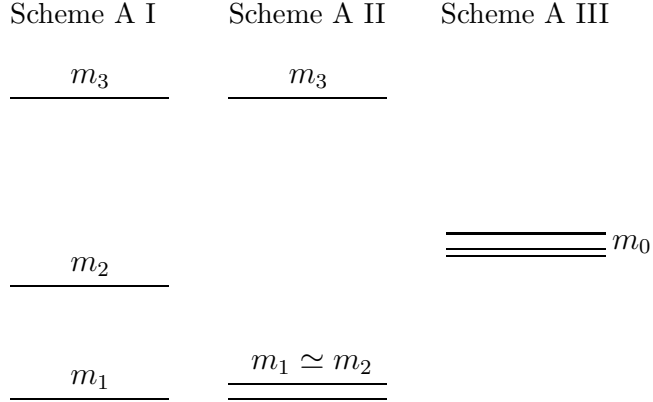

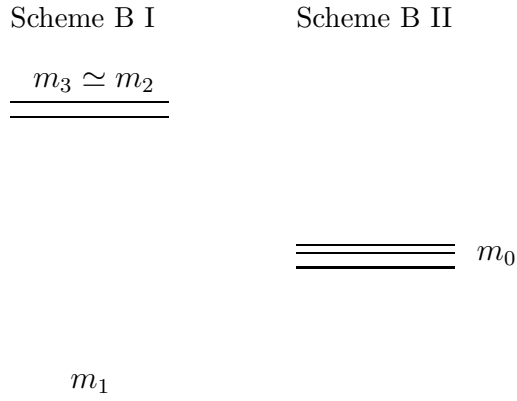
\begin{figure}[t]
\begin{center}
\begin{picture}(7,6)
\put(0,0){\line(1,0){2.1}}
\put(0,3.8){\line(1,0){2.1}}
\put(0,4){\line(1,0){2.1}}
\put(3.8,2){\line(1,0){2.1}}
\put(3.8,2.1){\line(1,0){2.1}}
\put(3.8,1.8){\line(1,0){2.1}}
\put(0.8,0.2){$m_1$}
\put(0.3,4.2){$m_3 \simeq m_2$}
\put(6.2,1.9){$m_0$}
\put(0,5){\bf {\small Scheme B I}}
\put(3.8,5){\bf {\small Scheme B II}}
\end{picture}
\vspace{1cm}
\caption{\label{massschemes2}The two mass schemes that can accommodate the 
relation $\Delta m^2_{32} \ll \Delta m^2_{21} \simeq \Delta m^2_{31}$.}
\end{center}
\end{figure}

The phenomenology of neutrino oscillations is well reviewed in the literature, 
see e.g.\ \cite{bggreport}. Flavor eigenstates $\nu_{\alpha}$ 
($\alpha = e ,\, \mu , \, \tau$) 
are connected to mass eigenstates $\nu_i$ ($i = 1,2,3$) 
via an unitary matrix, i.e.\ $\nu_{\alpha} = U_{\alpha i} \nu_i$. 
For Majorana neutrinos this matrix can be 
parametrized as \cite{Japaner2}
\be \label{Upara}
U_{\alpha i} = \left( \bad 
c_1 c_3 & s_1 c_3 e^{i \lambda_1} & s_3 e^{-i \delta} \\[0.2cm] 
(-s_1 c_2 - c_1 s_2 s_3 e^{i \delta}) e^{-i \lambda_1} 
& c_1 c_2 - s_1 s_2 s_3 e^{i \delta} 
& s_2 c_3 e^{i (\lambda_2 - \lambda_1)}\\[0.2cm] 
(s_1 s_2 - c_1 c_2 s_3 e^{i \delta}) e^{-i \lambda_2} & 
(- c_1 s_2 - s_1 c_2 s_3 e^{i \delta}) e^{-i (\lambda_2 - \lambda_1)} 
& c_2 c_3\\ 
               \ea   \right)          
\ee
with $c_i = \cos\theta_i$ and $s_i = \sin\theta_i$. Three CP--violating 
phases are present. 
For neutrino oscillations, only one phase ($\delta$) 
contributes \cite{onlyone}. Effects of all phases are discussed 
e.\ g.\ in \cite{Japaner2}, for our estimations 
we shall skip them, since we are not interested in CP violation. 
In addition, the probabilities we use depend 
only on the absolute values of the mixing matrix elements. 
The probability of a flavor state $\alpha$ to oscillate 
into a state $\beta$ is given by 
\be
P_{\alpha \beta} = \delta_{\alpha \beta} - 2 \mbox{ Re } \sum\limits_{j > i} 
U_{\alpha i} U_{\alpha j}^{\ast} U_{\beta i}^{\ast} U_{\beta j} 
(1 - \exp{i \Delta_{ji}})  . 
\ee
Here
\[
 \Delta_{ji} =  \frac{L}{2E} \, \Delta m_{ji}^2   
= 2.54 \frac{L/\rm km}{E/\rm GeV} \Delta m_{ji}^2 /{\rm eV^2} \mbox{ with } 
\Delta m_{ji}^2 = m_j^2 - m_i^2.
\] 
Without loss of generality we assume 
\be \label{m3m2m1} 
m_3 > m_2 > m_1 > 0  . 
\ee 
With the above relation there are two possibilities to accommodate  
$\Delta m_{\odot}^2 \ll \Delta m_{\rm A}^2 $: 
\be \label{scenarios}
\baz 
{\bf Scenario \, A:} &  \Delta m_{21}^2 = \Delta m_{\odot}^2 
\ll \Delta m_{\rm A}^2 = \Delta m_{31}^2 \simeq \Delta m_{32}^2  \\[0.2cm]
{\bf Scenario \, B:} &  \Delta m_{32}^2 = \Delta m_{\odot}^2 
\ll \Delta m_{\rm A}^2 = \Delta m_{21}^2 \simeq \Delta m_{31}^2  
\ea
\ee
Three mass schemes are capable of providing scenario A, 
qualitatively shown in Fig.\ \ref{massschemes1}: 
\be \label{EqschemesA}
\bad
\mbox{\bf Scheme A I:}  & m_3 \gg m_2 \gg m_1 \, : & 
m_3 \simeq \sqrt{\Delta m_{31}^2} \, , \; 
m_2 \simeq \sqrt{\Delta m_{21}^2} \\[0.2cm]
\mbox{\bf Scheme A II:}  & m_3 \gg m_2 \simeq m_1  \, : & 
m_3 \simeq \sqrt{\Delta m_{31}^2} \, , \; 
m_2 \simeq m_1 \simeq 0 \\[0.3cm]
\mbox{\bf  Scheme A III:}  & m_3 \simeq m_2 \simeq m_1 \equiv m_0  \, : & 
3 m_0 \simeq 5 \; \rm eV \, , \\[0.2cm]
\ea
\ee
where $m_0$ comes from cosmological considerations \cite{cosmo}.\\ 
\unitlength1cm
For scenario B, however, there are only two possibilities as 
presented in Fig.\ \ref{massschemes2}: 
\be \label{EqschemesB}
\bad
\mbox{\bf Scheme B I:}  & m_3 \simeq m_2 \gg m_1 \, : & 
m_3 \simeq m_2 \simeq \sqrt{\Delta m_{31}^2} \, , \; 
m_1 \simeq 0  \\[0.2cm]
\mbox{\bf  Scheme B II:}  & m_3 \simeq m_2 \simeq m_1 \equiv m_0  \, : & 
3 m_0 \simeq 5 \; \rm eV \, .  \\[0.2cm]
\ea
\ee
We stress that there are no other possibilities when 
Eqs.\ (\ref{scales}) and (\ref{m3m2m1}) are used. 
Scenario A I can be obtained from the see--saw mechanism \cite{seesaw}; 
A II and B I, i.e.\ two very close masses and one separated by the others 
can be a result of mechanisms generating neutrino masses radiatively 
\cite{zeebabu}. 
\unitlength1cm
Regardless of the concrete scheme, 
relations (\ref{scales}) and (\ref{Upara}) allow to get 
the absolute values of the 
elements of the MNS matrix \cite{decoupling}: 
In scenario A one finds for a 
short baseline reactor experiment 
such as CHOOZ: 
\be 
P_{ee}^{\rm CHOOZ} = 1 - 4 \ued (1 - \ued) \sin^2 \Delta_{31}/2 
\ee 
Due to the negative results CHOOZ presents \cite{CHOOZ}, 
one comes to the conclusion that 
\ued{} is either very small or close to 1. Taking into account that 
the probability $P_{ee}^{\odot}$ for solar neutrinos is significantly 
lower than 1, leads to a small value of \ued{} \cite{mee1}. 
In that case one has  
\be \label{solprob}
P_{ee}^{\odot} = 1 - 4 \uee \uez \sin^2 \Delta_{21}/2 . 
\ee
For oscillations of atmospheric 
$\nu_{\mu}$'s into $\nu_{\tau}$'s one finds 
\be \label{atmprob}
P_{\mu \tau}^{\rm A} = 4 \umd \utd \sin^2 \Delta_{31}/2   . 
\ee
In obtaining this equation we have assumed that the oscillation triggered by 
$\Delta m_{21}^2$ washes out. This is the case for 
$L({\rm km})/E(\rm GeV) \ll 10^5$ so that it is advisable to use 
the through--going muons data set of SuperKamiokande (SK). 
Equations (\ref{solprob}) and (\ref{atmprob}) together with the unitarity of 
the MNS matrix are now used to get the absolute values of 
all elements \cite{decoupling}. 
As the mass and mixing parameters we use the best fit points from 
\cite{wheredo} for solar neutrinos and for the SK through--going 
muon sample the values from \cite{soltoatm}.   
The numbers used are the following: 
\be \label{dataused}
\baz
(\Delta m_{21}^2 \; {\rm (eV^2)}, \; 4 \uee \uez ) = & 
\left\{ \baz ( 5.4 \cdot 10^{-6},6.0 \cdot 10^{-3}) & \mbox{ SAMSW}  \\[0.2cm]
            ( 1.8 \cdot 10^{-5},0.76)  & \mbox{ LAMSW}\\[0.2cm]
            ( 8.0 \cdot 10^{-11},0.75)  & \mbox{ VO}\\[0.3cm] 
\ea \right. \\ 
(\Delta m_{31}^2 \; {\rm (eV^2)}, \; 4 \umd \utd ) = & 
( 1.0 \cdot 10^{-2} ,\;  0.78) 
\ea
\ee
Note the unusual high value for the atmospheric mass scale. 
We give now the resulting mixing and mass matrices, 
starting with scenario A and commenting on scenario B later.

\subsection{The case $|U_{e3}| \neq 0$} 
The value $ \Delta m_{31}^2 = 1.0 \cdot 10^{-2} \, \rm eV^2$ corresponds in 
CHOOZ's exclusion plot of \cite{CHOOZ} to 
$\ued{} \; \ls \; 0.04$. Translating this in Eq.\ (\ref{Upara}) leads to 
$|s_3| \; \ls \; 0.19$ and 
$|c_3| \; \gs \; 0.98 \simeq 1$.  
This leads to the following mixing matrices:
\be \label{Umatrix1}
|U_{\alpha i}| \, \ls \, \left\{
\baz
\left( \bad 0.999 & 0.039 & 0.189 \\[0.2cm] 
                                   0.131 & 0.860 & 0.515 \\[0.2cm] 
                                   0.182 & 0.521 & 0.857 \\ 
               \ea   \right) & \mbox{ SAMSW}\\[1cm]
\left( \bad 0.863 & 0.505 & 0.189 \\[0.2cm] 
                                   0.517 & 0.789 & 0.515 \\[0.2cm] 
                                   0.400 & 0.527 & 0.857 \\ 
               \ea   \right) & \mbox{ LAMSW}\\[1cm]
\left( \bad 0.866 & 0.500 & 0.189 \\[0.2cm] 
                                   0.513 & 0.791 & 0.515 \\[0.2cm] 
                                   0.398 & 0.527 & 0.857 \\ 
               \ea   \right) & \mbox{ VO.}         \\[1cm]
\ea \right. 
\ee
Here we have used that $|U_{e1}| \ge |U_{e 2}|$ as it is necessary for the 
MSW solutions and --- inspired by the CKM matrix --- 
$|U_{\tau 3}| \ge |U_{\mu 3}|$. 
The possibility $|U_{e2}| \ge |U_{e 1}|$ is equivalent to an 
exchange of the first and second column, 
as $|U_{\mu 3}| \ge |U_{\tau 3}|$ is to 
an exchange of the second and third row. 
\subsection{The case $|U_{e3}| \simeq 0$}
The observed smallness of \mee{} as measured in \obb{} \cite{doublebeta} 
and the absence of $\nu_e$ mixing in atmospheric oscillations 
has lead many authors (see e.\ g.\ \cite{mee1,mee2,decoupling,ued0})  
to the assumption $|U_{e3}| \simeq 0$. In addition, it can also be 
related (together with bi--maximal mixing) with the observed flatness in 
$L/E$ of SK`s $e$--like events \cite{ahlu}. 
Since also the phases 
$\lambda_{1,2}$ in Eq.\ (\ref{Upara}) do not contribute 
to oscillations, this means 
that there is no observable CP violation in any oscillation experiment.  
The mixing matrix now reads: 
\be \label{Umatrix2}
|U_{\alpha i}| \, \ls \, \left\{
\baz
\left( \bad 0.999 & 0.039 & 0   \\[0.2cm] 
                                   0.033 & 0.856 & 0.515 \\[0.2cm] 
                                   0.020 & 0.515 & 0.857 \\ 
               \ea   \right) & \mbox{ SAMSW}\\[1cm]
\left( \bad 0.863 & 0.505 & 0     \\[0.2cm] 
                                   0.433 & 0.740 & 0.515 \\[0.2cm] 
                                   0.260 & 0.445 & 0.857 \\ 
               \ea   \right) & \mbox{ LAMSW}\\[1cm]
\left( \bad 0.866 & 0.500 & 0     \\[0.2cm] 
                                   0.429 & 0.742 & 0.515 \\[0.2cm] 
                                   0.258 & 0.446 & 0.857 \\ 
               \ea   \right) & \mbox{ VO.}         \\[1cm]
\ea \right. 
\ee
The matrices are of course very similar to the ones derived in 
\cite{decoupling} where the method was first presented. 

\subsection{Scenario B}
Scenario B can be easily obtained from scenario A via 
cyclic permutation of the columns of the mixing matrices. 
Hence, the cases to distinguish are 
$\uee{} \neq 0$, $\uee{} \simeq 0$, $\uez \ge (\le) \ued$ (the ``$\le$'' 
case only for the vacuum solution) and 
$\ume \ge (\le) \ute$.  
With the two scenarios A and B and the possibilities for 
ordering the mixing matrix 
elements we have a total of 80 different mass matrices, 48 for scenario A and 
32 for scenario B\@. 
From these 80 matrices, 48 are stemming from nondegenerate 
mass schemes. We will show that the 80 reduces to 57.  

\section{Results for \mab}
As shown, the different possibilities are equivalent to exchanges of rows or 
columns of the mixing matrices. The same holds for the resulting 
mass matrices. 
For example, the difference of the $|U_{\tau 3}| \ge |U_{\mu 3}|$ case 
and the $|U_{\tau 3}| \le |U_{\mu 3}|$ case translates into an exchange of 
$\langle m_{\alpha \tau} \rangle $ with $\langle m_{\alpha \mu} \rangle $ 
with $\alpha = e , \mu$. Replacing $|U_{e 1}| \ge |U_{e 2}|$ 
with $|U_{e 1}| \le |U_{e 2}|$ has no effect on \mab{} as long as 
$|U_{e 3}| \neq 0$ and generally in the degenerate schemes and in 
A II and B I. 
This, with the 
appropriate permutations mentioned above, holds for scenario B as well. 
Evidently, the degenerate schemes give the same result for 
\mab{} in both scenarios. 
Therefore from all 80 possible mass matrices only 57 survive, 
14 degenerate and 43 nondegenerate ones. 
We give now for the nondegenerate cases and 
for all three solar solutions the bounds of our results: 
\be  \label{massresult}
\mab{} \; \; \ls \left\{ \baz 
\left( 
\bad 
3.5 \cdot 10^{-6} \ldots 0.1 & 4.6 \cdot 10^{-5} \ldots 2.0 \cdot 10^{-2} 
& 4.6 \cdot 10^{-5} \ldots 2.0 \cdot 10^{-2} \\[0.2cm]
               & 2.7 \cdot 10^{-2} \ldots 7.6 \cdot 10^{-2} 
& 6.6 \cdot 10^{-3} \ldots 4.7 \cdot 10^{-2} \\[0.2cm]
               &                   & 2.7 \cdot 10^{-2} \ldots 7.6 \cdot 10^{-2} 
\ea \right) \rm eV  & \mbox{ SAMSW } \\[1cm]
\left(
\bad 
1.1 \cdot 10^{-3} \ldots 0.1 & 9.5 \cdot 10^{-4} \ldots 8.5 \cdot 10^{-2} 
& 9.5 \cdot 10^{-4} \ldots 8.5 \cdot 10^{-2} \\[0.2cm]
               & 2.7 \cdot 10^{-2} \ldots 8.9 \cdot 10^{-2} 
& 4.4 \cdot 10^{-2} \ldots 6.2 \cdot 10^{-2} \\[0.2cm]
               &                   & 2.7 \cdot 10^{-2} \ldots 8.9 \cdot 10^{-2}  
\ea \right) \rm eV  & \mbox{ LAMSW } \\[1cm]
\left(
\bad 
2.2 \cdot 10^{-6} \ldots 0.1 & 2.0 \cdot 10^{-6} \ldots 8.4 \cdot 10^{-2} 
& 2.0 \cdot 10^{-6} \ldots 8.4 \cdot 10^{-2} \\[0.2cm]
               & 2.7 \cdot 10^{-2} \ldots 8.9 \cdot 10^{-2} 
& 4.4 \cdot 10^{-2} \ldots 6.2 \cdot 10^{-2} \\[0.2cm]
               &             & 2.7 \cdot 10^{-2} \ldots 8.9 \cdot 10^{-2} 
\ea \right) \rm eV  & \mbox{ VO. } \\[1cm]
\ea  \right.
\ee
In the scheme (A II, $|U_{e 3}| \simeq 0$) there are zeros as solutions for 
$\langle m_{e \alpha} \rangle $, which means values much smaller than 
the atmospheric scheme, i.e.\ possibly in the range of 
$10^{-3} \ldots 10^{-4}$ eV, e.g.\   
\be
\langle m_{e e} \rangle \le \uee{} m_1 + \uez{} m_2 + \ued{} m_3 
= m_1 \ll \sqrt{\Delta m_{\rm A}^2} . 
\ee
The same can happen in scheme A I, where always a contribution of 
a value much smaller than the solar scheme can be present.  
These cases reflect the fact that the smallest mass eigenvalue is never 
known. 
For the degenerate scheme the solutions for VO and LAMSW are almost 
identical:  
\be  \label{massresultdeg}
\mab{} \; \; \ls \left\{ \baz 
\left( 
\bad 
1.67 \ldots 1.73  & 0.74 \ldots 1.57 & 0.74 \ldots 1.57  \\[0.2cm]
                  & 1.67 \ldots 1.93 & 1.47 \ldots 1.77  \\[0.2cm]
                  &                  & 1.67 \ldots 1.95 
\ea \right) \rm eV &  \mbox{LAMSW and VO} \\[1cm]
\left(
\bad
1.67 \ldots 1.73  & 0.07 \ldots 0.61 & 0.07 \ldots 0.61  \\[0.2cm]
                  & 1.67 \ldots 1.73 & 1.47 \ldots 1.52  \\[0.2cm]
                  &                  & 1.67 \ldots 1.73 
\ea \right) \rm eV &  \mbox{SAMSW.}
\ea \right.
\ee
The values $>m_0$ are explained by the violation of unitarity of the 
mixing matrices.

\subsection{Properties of the mass matrices}
We start with scenario A: 
In general, for the hierarchical schemes,  
\mmm{} is the biggest, \mee{} the smallest entry in \mab{}. 
The difference can be up to 4 orders of magnitude 
(VO and SAMSW, scheme A I, $\ued \simeq 0$). 
Entries in the electron row of \mab{} are smaller than the other elements.  
The degenerate scheme has always \mee{} $\simeq$ \mmm{} $\gs$ \meu{}. 
In addition, the $\ued \neq 0$ case delivers higher values for all 
elements of \mab{}. 
Scheme III gives higher numbers than scheme I which in turn gives higher 
values than scheme II\@.\\ 
In scenario B all entries are usually in the same order of magnitude yet 
somewhat higher than in scenario A. 
The element \mee{}, which is the most natural candidate for experimental 
access, is the only one which is significantly higher,  
at least one order of magnitude. It is always bounded by 
$\sqrt{\Delta m_{\rm A}^2}$, see the next section.  
As said before, the degenerate scheme gives the 
same numbers in both scenarios. 
Two typical matrices show most of the mentioned points, the first for 
scheme (A I, LAMSW, $|U_{e 1}| \ge |U_{e 2}|$, 
$|U_{\mu 3}| \ge |U_{\tau 3}|$ and $|U_{e 3}| = 0$) while the second 
is for (B II, VO, $|U_{e 2}| \ge |U_{e 3}|$, 
$|U_{\mu 1}| \ge |U_{\tau 1}|$ and $|U_{e 3}| \neq 0$):   
\be 
\baz 
\mab{} \; \; \ls 
 \left( \bad 3.5 \cdot 10^{-6} & 4.6 \cdot 10^{-5} 
                                      & 7.7 \cdot 10^{-5}\\[0.2cm]
                               &  7.4 \cdot 10^{-2} 
                                      & 4.5 \cdot 10^{-2} \\[0.2cm]
                               &           & 2.8 \cdot 10^{-2} 
\ea \right) \rm eV & \\[1cm]  
\mab{} \; \; \ls 
 \left( \bad 1.73 & 1.28  & 1.56 \\[0.2cm]
                  &  1.95 & 1.77 \\[0.2cm]
                  &       & 1.92
\ea \right) \rm eV. & \ea
\ee

\subsection{Electron neutrino mass}
From inspection of Eq.\ (\ref{massresult}) one sees that 
the only element accessible to present or near future experiments, \mee{}, 
has the broadest spectrum, a lucky coincidence. On the other hand, 
due to $\ued \ll 1$, \mee{} is always the smallest entry for scenario A\@.  
From all 57 matrices 33 different possibilities for \mee{} exist 
(it does not matter if $|U_{\mu 3}| \ge|U_{\tau 3}|$ or vice versa). 
From these 33 values \mee{} takes 
only 10 different values, which are worth taking a closer look at  
since they spread 6 orders of magnitude. Leaving the ones obtained from 
degenerated schemes aside, since they lie already above the current 
experimental limit, we have 8 different values out of 25 matrices, 
spanning 5 orders of magnitude. It is now tempting to assume that these 
8 values have potential to distinguish between the different solar solutions. 
This is unfortunately not the case:\\ 
As a result of relation (\ref{scales}) many schemes have the same value for 
\mee{}, for example  (A I, SAMSW, $|U_{e 1}| \ge |U_{e 2}|$  
and $|U_{e 3}| \neq 0$),  
(A II, all solar solutions, $|U_{e 1}| \ge |U_{e 2}|$ 
and $|U_{e 3}| \neq 0$) or 
(A I, VO, $|U_{e 2}| \ge |U_{e 1}|$  
and $|U_{e 3}| \neq 0$), all yielding $3.6 \cdot 10^{-3}$ eV\@.\\
In addition, for scenario B a 
peculiarity occurs, namely the bound always takes the same value, 
regardless of the solar solution:   
\be
\mee_{\rm B} \le \uee{} m_1 + \uez{} m_2 + \ued{} m_3 
\simeq m_3 \simeq \sqrt{ \Delta m_{\rm A}^2}   . 
\ee
In the nondegenerate schemes of scenario B, this value of \mee{} is 
always the largest entry in \mab{}. 
Hence a positive signal in a \onbb{} experiment will not be able to 
distinguish between different solar solutions, if nature has chosen scenario B 
for its neutrinos and an analysis of this kind is used. In addition, when 
scenario A is realized it will be extremely challenging to distinguish the 
precise form of the mass and mixing scheme as well as to tell which 
solar solution is the right one. What also complicates the analysis is 
that, as mentioned above, the contribution of the neglected $m_1$ to 
\mee{} can very well be in the order of $10^{-3}$ eV, making a definite 
statement somewhat difficult.\\
Nevertheless, scenario B is only a factor 2 from current experimental 
limits away, lying well within access in updates of the $^{76}$Ge experiment. 
Although we used a somewhat high value for $\Delta m_{\rm A}^2$ this 
specific situation seems to be the only realistic candidate for 
experimental detection. In the next section we will discuss 
the possibilities of detecting a process sensitive on 
\mab{} in more detail.

\section{\label{vier}Lepton--number and --flavor 
violating processes and Majorana neutrinos}
For this section it is important to stress again the difference between 
{\it direct} bounds, i.e.\ considering processes that depend on 
the respective matrix element and {\it indirect} bounds obtained in 
the present paper, i.e.\ using oscillation data and unitarity 
of the MNS matrix.\\  
As not surprising, \onbb{} is the best examined process triggered by 
Majorana neutrinos, resulting in a limit of 
$\mee \le 0.2$ eV \cite{doublebeta}. 
The electron--muon element \meu{} can be inferred from 
muon--positron conversion in sulfur nuclei. Theoretical estimations from 
\cite{doi} together with the PDG \cite{PDG} limit of the 
branching ratio give a limit of $\meu \le 0.4 \; (1.9)$ GeV, 
when the final state proton pairs are in spin singlet (triplet) 
state, respectively. 
The very same diagram as for \obb{} can be applied to other processes like 
$K^+ \ra \pi^- \mu^+ \mu^+$, which has an experimental 
branching ratio limit of 
$1.5 \cdot 10^{-4}$ \cite{littenberg}. 
Taking the calculation of \cite{abad}, one finds a limit of 
$\mmm \le 1.1 \cdot 10^{5}$ GeV\@. Another process depending on \mmm{} is 
$\mu^- \mu^+$ conversion in Titanium, discussed in \cite{moha}. 
Instead of nuclear captions or rare decays it was shown in \cite{FRZ1}, 
that in principle one can use high--energy scattering processes --- in this 
case tri--muon production at fixed target 
neutrino--nucleon experiments --- to get a bound on \mmm. 
Without worrying too much about experimental cuts a limit of 
$\mmm \, \ls \, 10^{4}$ GeV was obtained. 
In \cite{FRZ2} this procedure was applied to existing HERA data 
and generalized to the process 
\be \label{HERAprocess}
e^+ p \ra \aen \alpha^+ \beta^+ X \mbox{ with }  
(\alpha \beta) = (e \tau) , \; (\mu  \tau) , \;(\mu \mu) 
\mbox{ and } (\tau \tau)  , 
\ee
giving for the first time {\it direct} limits on the tau--sector of the 
mass matrix. Another direct way to obtain information about the 
tau sector of \mab{} might be $B^+ \ra X^- \tau^+ \alpha^+$ with 
$\alpha = e , \mu \mbox{ or } \tau$ and $X= \pi , K, D, \ldots$.   
In total, the current situation for bounds 
deduced from processes directly depending on \mab{} reads: 
\be \label{meffresult}
\mab{} \; \; \ls 
 \left( \bad 2 \cdot 10^{-10} & 0.4 \; (1.9) & 4.2 \cdot 10^{3}\\[0.2cm]
                              &  4.0 \cdot 10^{3} & 4.4 \cdot 10^{3} \\[0.2cm]
                              &           & 2.0 \cdot 10^{4} 
\ea \right) \rm GeV .  
\ee
A spread over 14 orders of magnitude can be seen. An improvement of the  
values is surely advisable. The somewhat unusual way to use 
high--energy scattering as done in \cite{FRZ1,FRZ2} is highly 
compatible with decay analyses: For example, assuming that a 
branching ratio for $ K^+ \ra \pi^- \mu^+ \mu^+$ of about 
$9.2 \cdot 10^{-8}$ (BR for $ K^+ \ra \pi^+ \mu^+ \mu^-$, \cite{Kdecay})  
can be achieved would result in a 
limit of $\mmm \; \ls \; 3.5 \cdot 10^3$ GeV, almost the same number 
as from HERA data, which itself will be improved by luminosity and 
energy updates.\\ 
There are other Majorana induced $\Delta L \neq 0$ processes, 
which are at present 
however not experimentally accessible: Running an $e^+ e^-$ collider 
in $e^- e^-$ mode could give rise to the ``inverse \onbb{}'' 
$e^- e^- \ra W^- W^- $ \cite{londontalk}. 
The same could be done for a $\mu \mu$ collider 
or even a possible $e \mu$ machine. 
For the case we are interested in ($s \gg m_i^2$) the cross section 
reads  
\cite{inv2beta} 
\be
\sigma (\alpha^- \beta^- \ra W^- W^-) \simeq  
\frac{G_F^2}{4 \pi} \mab{}^2 
\simeq 4.30 \cdot 10^{-17} \left( \frac{\mab{} }{\rm eV} \right)^2 
\mbox{ fb} \, , 
\ee
leaving no prospects for detection, since the \cs{} is in the order of 
$10^{-20}$ ($10^{-16}$) fb for hierarchical (degenerate) scheme.\\
For the sake of completeness one has to add a few words on 
cancellation of terms 
in \mab{}. From Eqs.\ (\ref{defin}) and (\ref{m3m2m1}) it 
is clear that our bounds are insensitive on the phases of the mixing 
elements. See \cite{Japaner2} for more details on what one could learn from 
the different phase dependence of \mee, \meu{} and \mmm.  
The Majorana nature brings additional complications via the 
intrinsic CP--parities  $\eta_i^{\rm CP}$  
of the mass eigenvalues. Even for CP invariance 
one can write \cite{bilpet}
\be
\langle m_{\alpha \alpha} \rangle = 
\left| \sum \left|U_{\alpha i} \right|^2 m_i \eta_i^{\rm CP} \right| . 
\ee  
making destructive interference in the respective amplitudes possible. 
Then one can assume mass matrices in flavor space which 
prohibite special entries in \mab{}, e.g.
\be
\mab{} = \left( \baz 0 & m \\[0.2cm]
                     m & 0 \ea \right)  \; , \, m \ge 0 . 
\ee 
with eigenvalues $\pm m$, 
first introduced to conserve $L_e - L_{\mu}$. Here, 
the requirement $m_i \ge 0$ can be saved by making the mixing 
matrix elements complex, leading again to cancellation.\ 
However, as long as no evidence for a nonvanishing 
element of \mab{} is found, we have no chance to decide which 
of the above possibilities is realized by nature. Yet, if in direct 
mass searches for, say, the $\nu_e$ a result of a few eV
is found, or the degenerate scheme is somehow verified,
one could give bounds on the phases and thus restrict different models. 

\section{Conclusions}
\begin{table}[ht]
\begin{center} 
\begin{tabular}{c|c|c|c|c|c}
Element & max.\ (eV) & Process & Ratio &  $\frac{\rm Theory}{\rm Data}$  
& $\frac{\rm direct}{\rm indirect}$ \\ \hline \hline
\mee{} & $ 0.1 $ & \nnu{}   
& $T_{1/2}^{0 \nu} \ge 2.3 \cdot 10^{26} $ y & 0.25 & 2.0  \\ \hline
\meu{} & $8.5 \cdot 10^{-2}$ 
& $^{32}{\rm S}(\mu^- , e^+) ^{32}{\rm Si}$ 
& $ \ba 4.5 \cdot 10^{-29} \\ 1.8 \cdot 10^{-30} \ea $ 
& $ \ba 5.0 \cdot 10^{-20} \\ 2.0 \cdot 10^{-21} \ea $ 
& $ \ba 4.7 \cdot 10^{9} \\ 2.2 \cdot 10^{10} \ea $
\\ \hline 
\met{} & $8.5 \cdot 10^{-2}$ & $e^+ p \ra \aen{} e^+ \tau^+ X$ 
& $ 3.2 \cdot 10^{-31} $ & $ 4.0 \cdot 10^{-28} $  & 
$4.9 \cdot 10^{13}$\\ \hline
\mmm{} & $2.6 \cdot 10^{-2}$ 
& $ \ba K^+ \ra \pi^- \mu^+ \mu^+ \\ 
e^+ p \ra \aen{} \mu^+ \mu^+ X \ea$ 
& $\ba 6.5 \cdot 10^{-35} \\ 3.8 \cdot 10^{-31} \ea $ & 
 $\ba 4.4 \cdot 10^{-31} \\ 4.9 \cdot 10^{-28} \ea $  
& $ 1.5 \cdot 10^{14} $ \\ \hline
\mmt{} & $6.2 \cdot 10^{-2}$ & $e^+ p \ra \aen{} \mu^+ \tau^+ X$ 
& $ 1.6 \cdot 10^{-31} $ & $ 2.0 \cdot 10^{-28} $  
& $7.1 \cdot 10^{13}$ \\ \hline
\mtta{} & $8.9 \cdot 10^{-2}$ & $e^+ p \ra \aen{} \tau^+ \tau^+ X$ 
& $ 1.5 \cdot 10^{-32} $ & $ 1.9 \cdot 10^{-29} $  
& $2.2 \cdot 10^{14}$ \\ 
\end{tabular}  
\caption{\label{tabhie}Element of the mass matrix in hierarchical schemes   
together with its maximal value, a process sensitive to the mass 
and a ratio with respect to the appropriate standard model process. 
Note that these highest values come only from scenario B I. 
Scenarios A I and II have typically mass values one or two orders 
of magnitude lower, thus the numbers in the 
last three columns are 2 to 4 orders worse.}
\end{center}
\end{table}  
Tables \ref{tabhie} (hierarchical scheme) 
and \ref{tabdeg} (degenerate) summarize our results 
for the different elements of the mass matrices, 
together with their maximal values obtained from our estimations, 
the process sensitive to the respective element, a ratio with respect 
to the relevant standard model process (see below) and 
a number which indicates how far away we are 
from detecting the process and thus having access to the element. 
Also given is the ratio of our 
indirect bound with the previous direct limits from Eq.\ (\ref{meffresult}).\\
As the ratio we use for the $K$ decay the branching ratio and for the   
HERA processes (\ref{HERAprocess}) the quotient of the  
respective \cs{} with the usual standard model charged current process of 
$\sigma (e^+ p \ra \aen{} X,\, Q^2 > 200 \; \rm GeV^2) \simeq 30.3$ 
pb \cite{HERAdata}. 
For \mee{} we give the half--life obtained with the matrix elements 
from \cite{nullbetamat} for $^{76} \rm Ge$.\\ 
Theory/Data is a measure for how close (better: how far away) 
we are from detection of the respective process. 
We use for \mee{} the current experimental limit for 
$T_{1/2}^{0 \nu} \, (^{76}\rm Ge)$ divided by our bound, 
for the $K$ decay our result divided by the current measured BR limit and 
for the HERA processes the \cs{} times the mean value of the 
luminosity analyzed by H1 \cite{isollepH1} and ZEUS \cite{isollepZEUS} in 
searches for isolated lepton events, 
$\mbox{${\cal L}$}_{e^+} = 42.1$ pb$^{-1}$. 
Note the almost hierarchical structure from $ee$ to $\tau\tau$ processes. 
Theory/Data is a number which characterizes how difficult it is
to investigate the respective effective mass. A value less than $10^{-3}$ 
to $10^{-4}$ for a given \mab{} cannot be regarded as accessible 
in laboratory experiments, even with very positive upgrade assumptions. 
The numbers show that no element other than \mee{} provides a  
realistic chance of accession. 
Regarding scenario A, the highest value obtained for \mee{} 
is $4.7 \cdot 10^{-3}$ eV, about the limit achievable 
in the most positive assumption of the 
GENIUS sensitivity of $2 \cdot 10^{-3}$ eV at 
68 $\%$ C.\ L.\  for a 10 year run with 10 tons of 
enriched germanium \cite{GENIUSproposal}. 
\begin{table}[ht]
\begin{center} 
\begin{tabular}{c|c|c|c|c|c}
Element & max.\ (eV) & Process & Ratio & $\frac{\rm Theory}{\rm Data}$  
& $\frac{\rm direct}{\rm indirect}$ \\ \hline \hline
\mee{} & $1.73 $ & \nnu{}   
& $T_{1/2}^{0 \nu} \ge 7.6 \cdot 10^{23} $ y 
& 75.7 & 8.7 \\ \hline
\meu{} & $1.57 $ 
& $^{32}{\rm S}(\mu^- , e^+) ^{32}{\rm Si}$ 
& $ \ba 1.5 \cdot 10^{-26}  \\ 6.1 \cdot 10^{-28} 
 \ea $ 
& $ \ba 1.7 \cdot 10^{-17}  
\\ 6.7 \cdot 10^{-19}  \ea $ 
& $ \ba 2.5 \cdot 10^{8}  \\ 1.2 \cdot 10^{9} 
 \ea $ \\ \hline 
\met{} & $1.57  $ & $e^+ p \ra \aen{} e^+ \tau^+ X$ 
& $ 1.1 \cdot 10^{-28} $ 
& $ 1.4 \cdot 10^{-25} $ & $2.7 \cdot 10^{12}$ \\ \hline
\mmm{} & $1.93 $ 
& $ \ba K^+ \ra \pi^- \mu^+ \mu^+ \\ 
e^+ p \ra \aen{} \mu^+ \mu^+ X \ea$ 
& $\ba 3.0 \cdot 10^{-32}  \\ 1.8 \cdot 10^{-28} 
 \ea $ & 
 $\ba 2.0 \cdot 10^{-28}  \\ 2.2 \cdot 10^{-25} 
 \ea $ & $7.7 \cdot 10^{12}$ \\ \hline
\mmt{} & $1.77 $ & $e^+ p \ra \aen{} \mu^+ \tau^+ X$ 
& $ 1.3 \cdot 10^{-28}  $ 
& $ 1.6 \cdot 10^{-25}  $ & $2.5 \cdot 10^{12}$ \\ \hline
\mtta{} & $1.95 $ & $e^+ p \ra \aen{} \tau^+ \tau^+ X$ 
& $ 6.9 \cdot 10^{-30} $ & $ 8.6 \cdot 10^{-27} 
 $ & $1.0 \cdot 10^{13}$ \\ 
\end{tabular}  
\caption{\label{tabdeg}Same as previous table for the mass limits obtained 
from the degenerate mass scheme. Note that the highest values come from 
the LAMSW or VO solution respectively.}
\end{center}
\end{table}  
Note however that we used the best--fit points of 
typical oscillation experiment analyses, which of course do not need to 
be the final answer. 
However, our value $\Delta m_{\rm A}^2$ = 0.01 eV$^2$, the best--fit 
point for through--going muons at SK, is just the maximum of 
typical general analyses (see \cite{soltoatm} for details) 
and therefore the values can be regarded 
as a realistic indication.  
In scenario B \mee{} is bounded by the atmospheric mass scale, 
thus lying well within the range of next generation 
\onbb{} experiments. A detection of a \mee{} in the range of the 
atmospheric scale would rule out the hierarchical schemes in scenario A.\\
Direct/indirect shows the ratio of the indirect limits obtained here and 
the direct limits from Eq.\ (\ref{meffresult}). This number can be as high 
as $10^{14}$. The indirect bounds obtained in this paper are more 
stringent by this number.\\ 
The tables show that the clarification of the question whether 
neutrinos are Dirac or Majorana particles might have to be postponed. Present 
and foreseeable experimental possibilities are far beyond verifying some 
of the given branching ratios or cross sections. This is 
shown by the ratios in the Theory/Data column. 
In conclusion, we used a three neutrino framework and studied the 
range of the elements of the effective mass matrix using all 
possibilities for the mass and mixing schemes. 
It turned out that \mee{} has the broadest range of all elements but 
also is in scenario A the smallest entry and in the nondegenerate schemes 
of scenario B the highest entry. 
in general is the smallest value of \mab{}. 
We may summarize the situation in saying that if scenario A is 
realized little hope should one have, whereas scenario B 
provides a realistic chance of being probed, leaving 
\onbb{}, the ``gold--plated process'' 
of lepton--number/flavor violation. The nice possibility of 
deciding which solution to the solar neutrino problem is realized 
is unfortunately not given, but this question will be 
clarified anyway in ongoing and forthcoming oscillation experiments.   
For a degenerate scheme the bounds are 
higher than current experimental limits so that a cancellation of 
terms in \mee{} as discussed at the end of Section \ref{vier} 
has to occur, provided nature has chosen this scheme.  
Regarding the other elements of \mab{} we showed that there is no 
possibility to investigate processes depending on them. 
Nevertheless, the improvement with 
respect to the previous direct bounds is 
up to 14 orders of magnitude.\\
{\it Note added:} When this paper was finished, Ref.\ \cite{kaiplb} 
appeared which gives new limits on two elements of \mab{}.

\newpage
\begin{center}
{\bf {\large Acknowledgments}}
\end{center}
I thank S.\ M.\ Bilenky, M.\ Flanz, E.\ A.\ Paschos 
and K.\ Zuber for helpful discussions. 
This work has been supported in part by the
``Bundesministerium f\"ur Bildung, Wissenschaft, Forschung und 
Technologie'', Bonn under contract number 05HT9PEA5.  
Financial support from the Graduate College 
``Erzeugung und Zerf$\ddot{\rm a}$lle von Elementarteilchen'' 
at Dortmund university is gratefully acknowledged.

\end{document}